\documentclass[vecphys]{svmult}
\usepackage{amsmath}
\usepackage{amssymb}

\usepackage{makeidx}         % allows index generation
\usepackage{graphicx}        % standard LaTeX graphics tool
\usepackage{multicol}        % used for the two-column index
\usepackage{cite}            % adjusts the "syntax" of the refs in the
\usepackage[hypertex]{hyperref}                             % text
\usepackage[bottom]{footmisc}% places footnotes at page bottom

\DeclareMathOperator{\Imm}{Im}
\DeclareMathOperator{\tr}{tr}

\DeclareMathOperator{\Rem}{Re}

\makeindex             % used for the subject index
\begin{document}

\title{Low temperature transport in tunnel junction arrays: Cascade energy relaxation}
\titlerunning{Low temperature transport in tunnel junction arrays}
% your contribution title if the original one is too long
\author{N.\,M.\,Chtchelkatchev \and V.\,M.\,Vinokur \and T.\,I.\,Baturina}
% Use \authorrunning{Short Title} for an abbreviated version of
% your contribution title if the original one is too long
\institute{Institute for High Pressure Physics, Russian Academy of Sciences,
Troitsk 142190, Moscow region, Russia
\texttt{nms@itp.ac.ru}
\and Materials Science Division, Argonne National Laboratory, Argonne, Illinois 60439, USA
\texttt{vinokour@anl.gov}
\and Institute of Semiconductor Physics, 13 Lavrentjev Avenue, Novosibirsk 630090, Russia
\texttt{tatbat@isp.nsc.ru}
}
% Use the package "url.sty" to avoid problems with special characters
% used in your e-mail or web address.
% Addresses should be removed from contribution and entered into
% blist.tex" (by the compiler).

\maketitle

\begin{abstract}
A theory of far-from-equilibrium transport in arrays of tunnel junctions is developed. 
We show that at low temperatures the energy relaxation ensuring  
tunneling current can become a cascade two-stage process. 
First, charge carriers 
lose their energy to a bosonic environment via non-phonon energy exchange.
The role of such an environment can be taken by 
electromagnetic fluctuations or dipole excitations (electron-hole pairs). 
The environment, in its turn, relaxes the energy to the thermostat by means 
of phonon irradiation.  
We derive the current-voltage characteristics for the arrays and demonstrate 
that opening the energy gap in the spectrum of the environmental excitations  
completely suppresses the tunneling current.  
The consequences of the cascade relaxation in various physical systems are discussed.

\end{abstract}

\noindent

\section{Introduction}\label{Sec:Intro}

Electronic transport in mesoscopic tunnel junctions is ensured
by the energy exchange between the tunneling charge carriers and energy reservoirs:
since the electronic energy levels at the banks of the mesoscopic junctions are, in general,
unequal, the tunneling is impossible unless there is a subsystem of excitations
capable of accommodating this
energy difference~\cite{Nazarov1989,Nazarov2007,Devoret_basic,Averin1990,Girvin1990,Ingold1991,Ingold-Nazarov,Ingold1994}.
Intense studies of nano-structured and disordered systems including
Josephson junctions~\cite{Aprili2009a,Aprili2009b},
mesoscopic tunnel junctions~\cite{STM1D} and superconductors~\cite{Timofeev2009},
patterned superconducting films~\cite{BaturinaSNSa,BaturinaSNSb},
highly disordered superconducting
and semiconducting films~\cite{Shahar2005,Baturina2007,Baturina2008a,Baturina2008b,Shahar2009,other}
reveal a prime importance of the \textit{out-of-equilibrium} properties of an environment to
which the tunneling charge carriers relax the energy. In particular, experiments on disordered 
superconducting films revealed that at extremely low temperatures the $I$-$V$ characteristics 
exhibit highly nonlinear behavior indicating that the transport charge carriers are decoupled  from the 
phonon thermostat.  

Notably, the relaxation processes can occur via relaxation of energy from tunneling carriers to some 
other bosonic environment mediated the energy exchange between the tunneling current and phonon thermostat.
The role of such bosonic mediator can be taken by ether the
electromagnetic environment~\cite{Nazarov1989,Nazarov2007,Girvin1990,Averin1990,Ingold1991,Ingold-Nazarov,Ingold1994,Kopnin_book,Giazotto}
by the neutral dipole excitations [the electron-hole (e-h) pairs] generated by the 
tunneling carriers~\cite{Lopatin1,Lopatin2}.  
By consideration in the spirit of Feynman-Vernon influence functional 
it was shown ~\cite{FVB,FVBlp,VinNature} that in the large Josephson junction arrays
the low temperature transport is governed by 
the unbinding neutral dipole excitations and the successive tunneling of the resulting electric
charges.
In the course of tunneling the propagating charges 
generate the dipole environment controlling relaxation process.
The formation of a gap in the dipole excitation spectrum impedes relaxation and causes
the suppression of the tunneling current, i.e. the localization of charge carriers. 

The energy relaxation in mesoscopic tunnel junctions in the case where
the energy exchange between the tunneling carriers and the electromagnetic
and/or electron-hole reservoir, $1/\tau_{\rm{e-env}}$,
is comparable to the rate of the energy loss to the phonon thermostat, $1/\tau_{\rm env\to bath}$,
was analyzed in~\cite{Kopnin2009}. In this article expanding on our kinetic approach 
developed in~\cite{CVB_PRL,CVB_PhysicaC} we
present a general approach to the description of strongly nonequilibrium
processes where $1/\tau_{\rm{e-env}}\gg 1/\tau_{\rm env\to bath}$
and show that the low-temperature energy relaxation enabling the tunneling current occurs in two stages:
(i) The energy relaxation from the tunneling charges
to the intermediate bosonic modes, electromagnetic or dipole (electron-hole) excitations,
which we hereafter call the \textit{environment}; and (ii)\,The energy transfer
from the environment to the phonon thermostat, to which we will be further referring as to
a \textit{bath}.  We demonstrate that at $1/\tau_{\rm{e-env}}\gg 1/\tau_{\rm env\to bath}$,
electronic transport is controlled by the first stage and is thus
critically sensitive to the spectrum of the environmental modes.  At the same time,
the passing current drives the environment out of the equilibrium, and
the environment spectrum and effective temperature may become bias-dependent themselves.
We construct coupled kinetics equations for
charge carriers and out-of-equilibrium bosonic environment and derive $I$-$V$ characteristics
for arrays of normal and superconducting junctions.

The article is organized as follows.
In Section 1.2 we construct a general theory of environment-mediated transport
in a single tunnel junction and calculate the corresponding $I$-$V$ characteristics.
In Section 1.3, we extend our theory onto arrays of tunnel junctions;
the content of these sections follows our recent
publications~\cite{CVB_PRL,CVB_PhysicaC}.
In Section 1.4 we explore application of the cascade two-stage energy
relaxation mediated transport to different physical systems and phenomena.
In particular, we discuss nonequilibrium Coulomb anomaly, negative differential resistance
and overheating in single junctions, properties of electron-hole environment in disordered
metals, and the mechanism of tunneling transport in a superinsulating state re-deriving the results 
of~\cite{FVB,FVBlp,VinNature} from a different perspective.

\section{A single junction }

\begin{figure}[b]
\includegraphics[width=1\columnwidth]{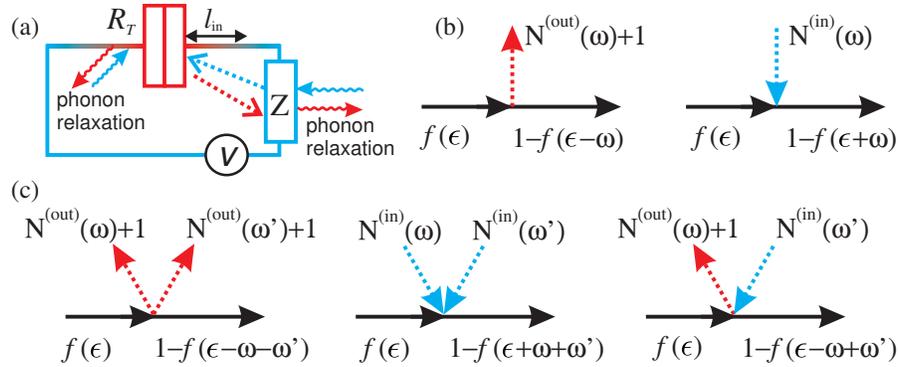}
\caption{
(a) The effective circuit for a tunnel junction subject to bias
$V$ and with the environment having the impedance $Z$.
(b)-(c) Diagrammatic expansion of $P^<$ to the first and the second orders in $\rho$, respectively.
The solid lines represent propagation of electrons,
the dashed lines denote the environment excitations.
The vertex with the two electron lines and one  dashed line
carries a factor $G_{\scriptscriptstyle{\mathrm T}}\rho(\omega)/\omega$, the
two dashed-lines vertex corresponds to
$G_{\scriptscriptstyle{\mathrm T}}\rho(\omega)\rho(\omega')/(\omega\omega')$.}
\label{fig:e-ph_int}
\end{figure}

We start with a tunnel junction between two bulk metallic electrodes biased
by the external voltage $V$, see Fig.\,\ref{fig:e-ph_int}a.
A general formula for the tunneling current reads:
          \begin{gather}\label{eq:current}
                  I=e\left(\overrightarrow{\Gamma} - \overleftarrow{\Gamma}\right)\, ,
           \end{gather}
where $\overrightarrow{\Gamma}$ ($\overleftarrow{\Gamma}$)
is the tunneling rate from the left (right) to the right (left), and, for a single junction,
   \begin{gather}\label{eq:S_begin_total}
        \overrightarrow{\Gamma}=\frac1{R_{\scriptscriptstyle {\mathrm T}}}
        \int_{\epsilon\epsilon'}f_\epsilon^{(1)}
        (1-f_{\epsilon'}^{(2)})P^<(\epsilon-\epsilon')\, ,
   \end{gather}
where $f^{(1,2)}$ are the electronic distribution functions within the
electrodes, $P^<(\epsilon)$
is the probability for the charge carrier to lose the energy $\epsilon$ to the environment, and
$R_{\scriptscriptstyle{\mathrm T}}$ is the bare tunnel resistance, representing the interaction of electrons with the bath.
The backward scattering
rate, $\overleftarrow{\Gamma}\propto\int_{\epsilon\epsilon'}f_\epsilon^{(2)}
(1-f_{\epsilon'}^{(1)})P^<(\epsilon-\epsilon')$. If  an intermediate environment is
absent and the relaxation is provided by the phonon bath, then
$P^<(\epsilon)=\delta(\epsilon)$ and  Eq.~\eqref{eq:S_begin_total}
reproduces the conventional Ohm law.
The quasiequilibrium situation
where the distribution functions of the environmental modes $N_{\omega}$
are Bose distributions parameterized
by the equilibrium temperature was discussed in~\cite{Devoret_basic,Ingold1991,Grabert_Devoret}.
In a general, far from the equilibrium case, we find:
%%%%%%%%%%%%%%%%%%%%%%%%%%%%%%%%%%%%%%%%%%%%%%%%%%%%%%%%%%
%%%%%%%%%%%%%%%%%%%%%%%%%%%%%%%%%%%%%%%%%%%%%%%%%%%%%%%%%%
\begin{gather}\label{eq:S_begin_total1}
   P^{<}(\epsilon)=\int_{-\infty}^\infty dt \exp[J(t)+i\epsilon t]\,,
  \\ \label{eq:J} J(t)=2\int_{0}^\infty \frac{d\omega}{\omega}\rho(\omega) F(\omega)\,,
    \\\notag
    F(\omega)=\left[N_\omega e^{i\omega t}+(1+N_\omega)e^{-i\omega t}-B_\omega\right]\,.
\end{gather}
%%%%%%%%%%%%%%%%%%%%%%%%%%%%%%%%%%%%%%%%%%%%%%%%%%%%%%%%%%
%%%%%%%%%%%%%%%%%%%%%%%%%%%%%%%%%%%%%%%%%%%%%%%%%%%%%%%%%%
Here $\exp[J(t)]$ is the nonequilibrium generalization of the Feynman-Vernon influence
functional~\cite{Feynman_Vernon} reflecting that tunneling electrons acquire random
phases due to interactions with the environment, represented by a set of
oscillators with the nonequilibrium distribution of modes, $N_{\omega}$.
The latter is defined by the kinetic equation with the scattering
integral describing the energy exchange between environmental modes and tunneling
electrons.
Terms proportional to the $N_\omega$ and $1+N_\omega$
correspond to the absorbed and emitted environmental excitations, respectively.
The combination $B_\omega=1+2N_\omega$
is the kernel of the time-independent contribution to $J$ describing the elastic
interaction of the tunneling electron with the environmental modes
and having the structure of the Debye-Waller factor.
In an equilibrium, $N_\omega$ reduces to the
Bose-function and the functional $P^<$ recovers the result by
Ref.~\cite{Grabert_Devoret}.
The spectral probability of the electron--(electromagnetic) environment interaction is
$\rho(\omega)=
\Rem[Z_{\mathrm t}(\omega)]/R_{\scriptscriptstyle{\mathrm Q}}$,
where $Z_{\mathrm t}=1/[iC\omega + Z(\omega)^{-1}]$ is the total circuit impedance,
$Z$ is the environment impedance, $C$ is the junction capacitance,
and $R_{\scriptscriptstyle{\mathrm Q}}$ is the  quantum resistance~\cite{Ingold1991}.
Proceeding analogously to
Ref.~\cite{Grabert_PRL}, one finds the spectral probability corresponding
to the electron--environment interaction within each electrode as
$\rho_{\mathrm n}(\omega)
=2\Imm\int_{\mathbf q}\tilde U_{\mathrm n}/(D_{\mathrm n} q^2-i\omega)^{2}$, $n=1,2$,
 and that for the interaction across the junction,
$\rho_{12}(\omega)=-2\Imm\int_{\mathbf q}\tilde U_{12}/
[(D_1q^2-i\omega)(D_2q^2-i\omega)]$, where $D_{1(2)}$
are diffusion coefficients within respective electrodes, and $\tilde U_{1(2)}$
are the dynamically screened Coulomb interactions within (across) the electrodes.

The spectral probability $\rho(\omega)$ of the electron--environment interaction
controlling the behavior of the $I$-$V$ characteristics is central to our approach.
The form of $\rho(\omega)$, in its turn, is determined by the structure of the environmental
excitation spectrum and, in general, depends on the external bias.
The latter dependence becomes essential
in the array of highly transparent junctions where $\rho(\omega)$ is different for elastic and
inelastic processes~\cite{Nazarov1989,Nazarov2007,SVB}.
In particular, for the e-h environment with constant $\tilde U$,
one should cut off the (diverging) integral at
$q_{\scriptscriptstyle{\mathrm T}}=\sqrt{1/\tau_{\varphi}(T_{\rm eff})D}$~\cite{footnote111},
when calculating $\rho(0)$. Here $\tau_{\varphi}$ is the electron inelastic time
and $T_{\rm eff}$ is
the (bias dependent) effective temperature of the environment defined below.
This allows us to formulate a recipe: if in an equilibrium $\rho=\rho(\omega,T)$ then in an
out-of-equilibrium state $\rho=\rho(\omega,T_{\rm eff})$.
Of special importance are the effects of spectrum reconstruction accompanying
the phase transformation in the system: it is the sensitivity of the shape of the
$I$-$V$ curves to the form of  $\rho=\rho(\omega,T)$ that makes tunneling currents an
irreplaceable spectroscopy tool.  In particular, opening the gap $T^*$ in the
environmental excitation spectrum suppresses $\rho=\rho(\omega)$ within the
energy interval $0<\omega<T^*$ giving rise to complete vanishing of the tunneling transport
current at low temperatures $T<T^*$.  As we will discuss below, this is
the microscopic mechanism behind the low-temperature suppression of electronic transport in
disordered arrays of tunnel junctions, the phenomenon of \textit{superinsulation}.

To close the set of formulas (\ref{eq:current})-(\ref{eq:S_begin_total1})
one has to add the kinetic equations (KE) for the boson distribution functions $N_{\omega}$.
To derive these KE we use a semi-phenomenological kinetic approach of~\cite{Landauvshiz_10}
and express the current of Eq.\,(\ref{eq:current}) through the
electronic distribution function as $I=\int_{\epsilon_1}[d  f^{(1)}_{\epsilon_1}/dt]\nu_1$.
Here $\nu_{1(2)}$ is the density of states in the lead $1(2)$ and
$df^{(1)}_{\epsilon_1}/{dt}=I_{\rm col}$,
where $I_{\rm col}$ is the collision integral
describing the evolution of the electronic distribution function due to
energy and/or momentum transfer processes.
Expanding further $P^<$ with respect to $\rho$ we obtain, in the zero order in $N_{\omega}$,
the collision integral in a form
$I_{\rm col}^{(0)}=
-\int W_{12}[f^{(1)}_{\epsilon_1}(1-f^{(2)}_{\epsilon_2})-
f^{(2)}_{\epsilon_2}(1-f^{(1)}_{\epsilon_1})]
\delta(\epsilon_1-\epsilon_2)\nu_2 d\epsilon_2$,
where $W_{12}=1/\nu_1\nu_2R_{\scriptscriptstyle {\mathrm T}}$
is proportional to the
bare probability for an electron to be transmitted from one lead to the other.
In the first order
\begin{gather}\label{eq:sc}
\begin{split}
\frac{df^{(1)}_{\epsilon_1}}{dt}=-&\int d\omega\nu_\omega \nu_2d\epsilon_2\left(\frac{\rho}{\omega\nu_\omega}\right)W_{12}\times\\
\biggl\{
&\delta(\epsilon_{12}-\omega)[f^{(1)}_{\epsilon_1}(N_\omega+1)
(1-f^{(2)}_{\epsilon_2})-(1-f^{(1)}_{\epsilon_1})N_\omega f^{(2)}_{\epsilon_2}]+
\\
&\delta(\epsilon_{12}+\omega)[f^{(1)}_{\epsilon_1}N_\omega
(1-f^{(2)}_{\epsilon_2})-(1-f^{(1)}_{\epsilon_1})
(N_\omega+1) f^{(2)}_{\epsilon_2}]\biggr\},
\end{split}
\end{gather}
where $\epsilon_{12}=\epsilon_1-\epsilon_2$ and $\nu_\omega$
is the density of environmental states \cite{env_DoS}.
The structure of
$I_{\rm col}^{(1)}$ is identical to that of the electron-phonon scattering integral in
metals \cite{Landauvshiz_10}, where $N_{\omega}$ would stand for
the phonon distribution functions.
The quantity $\rho/(\omega\nu_\omega)$ is proportional to the
probability of the electron-environment scattering.

The collision integral dual to $I_{\rm col}^{(1)}$
and describing the evolution of $N_{\omega}$ is derived analogously, and the
resulting kinetic equation is:
\begin{gather}\label{eq:N_e-env}
\left(\frac{dN_\omega}{dt}\right)_{\rm e-env}=
-\frac{A\rho(\omega)}{\nu_\omega R_{\scriptscriptstyle{\mathrm T}}}
\left[N_\omega (1+n_\omega)-(1+N_\omega)n_\omega\right],
\end{gather}
where $A$ is the numerical factor of order of unity, $n_\omega$
is the electron-hole pairs distribution function.
The scattering integral in Eq.\,\eqref{eq:N_e-env} is also identical by its structure to the
phonon-electron scattering integral in metals \cite{Landauvshiz_10}.
For the electron-hole environment ($i=1,2$ labels the electrode in which the pair is located),
one has
$n_\omega^{(i)}=
(1/\omega)\int_\epsilon f^{(i)}_{\epsilon_+}(1-f^{(i)}_{\epsilon_-})$,
where $\epsilon_\pm=\epsilon\pm\omega/2$;
this agrees with the results of Ref.~\cite{Kamenev-Andreev} where the nonequilibrium
bosonic distribution function is equivalent to our $1+2n_\omega$.
If electrons and holes belong to different electrodes,
\begin{gather}\label{eq:n}
n_\omega^{(ij)}=
(2\omega)^{-1}\int_\epsilon f^{(i)}_{\epsilon_+}\sigma^x_{ij}(1-f^{(j)}_{\epsilon_-}),
\end{gather}
where $\hat\sigma^x$ is the Pauli-matrix.
From Eq.\,\eqref{eq:N_e-env} one estimates the rate of the energy exchange
between the environment and
the tunneling electrons as:
$1/\tau_{\rm{e-env}}=\rho(\omega)/
(\nu_\omega R_{\scriptscriptstyle{\mathrm T}})$.
Now one has to compare
$1/\tau_{\rm{e-env}}$ with
the rate of the interaction of the environment modes with the (phonon) bath,
$1/\tau_{\rm env\to bath}(\omega)$.
For the electron-hole environment, $1/\tau_{\rm env\to bath}(\omega)$
is determined from Eq.\,\eqref{eq:N_e-env}
to which the electron-phonon scattering integral is added.
If $\tau_{\rm env\to bath}\gg \tau_{\rm e-env}$,  the two-stage energy relaxation takes place
and the characteristic energy transfer from tunneling current is
$\omega\sim \max\{T_{\mathrm e},V\}$, where
$T_{\mathrm e}$ is the electronic temperature in the leads.
The electromagnetic environment mediates the two-stage relaxation in the case where Ohmic
losses occur in a LC superconducting line and are small~\cite{env_DoS}.
To take a typical example, in aluminum mesoscopic samples $\tau_{\rm e-env}=10^{-8}$\,sec
and $\tau_{\rm env\to bath}=10^{-6}$\,sec~\cite{Giazotto},
so the conditions for the two-stage energy relaxation are realized.
Then the distribution functions, $N_\omega$, deviate significantly from
the Bose-distribution with the temperature of the phonon bath.  They should be
determined from the condition that the collision integral
of the environmental modes with the e-h pairs accompanying
the current flow becomes zero, and Eq.\,\eqref{eq:N_e-env}
yields $N_\omega\cong n_\omega^{(12)}$.
If $T_{\mathrm e}\ll V$, then $N_\omega$ can
be approximated by the Bose-function with some effective
temperature $T_{\rm eff}$ at $\omega<V=T_{\rm eff}$ and $N_\omega=0$ at $\omega>T_{\rm eff}$
(the emission of the excitations with  the energy larger than $V$ is forbidden), and
\begin{gather}\label{eq:T_eff}
T_{\rm eff}\equiv\lim_{\omega\to 0} \omega N_\omega=0.5 V\coth(V/2T)\,.
\end{gather}
This result shows that the system with the environment well isolated
from the bath cannot be cooled below $T_{\rm eff}$.
Note that the $\coth$-expression for $T_{\rm eff}$
obtained in the first approximation in $\rho$.
In a general case, $T_{\rm eff}$ depends on $\rho$.

Equations \eqref{eq:current}-\eqref{eq:N_e-env} give the full
description of the kinetics of the tunnel junction in a
nonequilibrium environment.  To derive the
$I$-$V$ characteristics we find $N_\omega\cong n_\omega^{(12)}$
and plug it into Eqs.\,\eqref{eq:current}-\eqref{eq:J}.
Introducing the parameters $g^{-1}=\rho(0)$ and $\Lambda$,
the characteristic frequency of the
$\rho(\omega)$ decay [for the
Ohmic model~\cite{Grabert_Devoret}, $\rho=g^{-1}/\{1+(\omega/\Lambda)^2\}$ and
$\Lambda/g$ is of the order of the charging energy of the tunnel junction], we find:
\begin{gather}\label{eq:I_g}
I\sim \frac V {R_{\scriptscriptstyle{\mathrm T}}}\ln \frac \Lambda V\,;
\end{gather}
in the interval $T\ll V\ll \Lambda$, where $T_{\rm eff}\simeq V$.
Note that $I(V)$ given by Eq.\,\eqref{eq:I_g} differs from the power law dependence
obtained in~\cite{Grabert_Devoret} for $T_{\mathrm e}=T_{\rm eff}=0$.
This shows that tuning  the environment
one can control the $I(V)$-characteristics of the tunnel junction (the gating effect).
At high voltages, $V\gg\Lambda$, one finds
\begin{gather}
    I(V)\simeq(V-\Delta_\infty)/{R_{\scriptscriptstyle{\mathrm T}}}\,,
    \\
    \Delta_\infty=iJ'(0)=2\int_0^\infty d\omega\rho_{\omega}[1+N_\omega^{\rm (out)}-N_{\omega}^{\rm (in)}]\simeq\Delta_\infty^{(0)}\ln(\Lambda/\min\{T_e,T_{\rm env}\}),
  \end{gather}
where $\Delta_\infty^{(0)}=\Delta_\infty[N^{\rm (out)}= N^{\rm (in)}]\sim \Lambda/g$,
since at $V\gg\Lambda$, $N^{\rm (out)}_{\omega}\simeq \Lambda/\omega\gg N^{\rm (in)}_{\omega}$.

\section{Arrays of tunnel junctions}

\begin{figure}[t]\center
  \includegraphics[width=0.8\columnwidth]{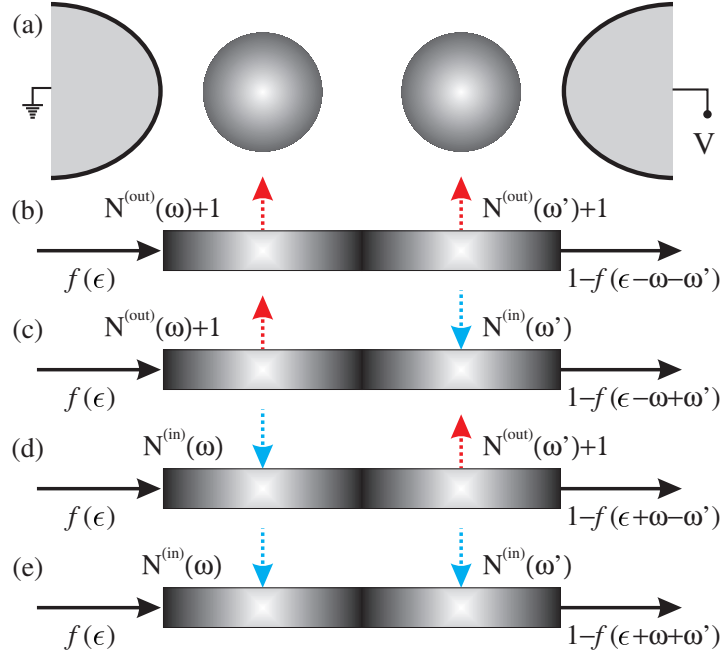}
  \caption{(a) The single electron two-islands' circuit.
(b)-(e) Diagrams describing the forward inelastic cotunneling rate.
The  ``up" arrows stand for the e-h pairs excited during the cotunneling
and the ``down" arrows correspond to the recombination of the e-h pairs.
The vertices shown by boxes are proportional to the probability
of an elemental e-h pair excitation, $\rho(\omega)/\omega$.
}
\label{fig:cotunneling}
\end{figure}

Extending Eq.\,(\ref{eq:S_begin_total}) onto an array comprised of $N$ junctions one finds
\begin{gather}\label{eq:cotunneling_rate}
    \overrightarrow{\Gamma}=\left(\prod_{i=1}^N\frac{R_Q}{4\pi^2 R_{i}}\right)\,S^2\,\int d\epsilon d\epsilon' f_1(\epsilon)[1-f_2(\epsilon')] P(\epsilon-\epsilon'),
\end{gather}
where
\begin{multline}
    P(E)= \int_{-\infty}^\infty dt \exp(iEt)
    \bigg\{\int_{0}^\infty d\omega\frac{\rho(\omega)}\omega \times%\right.
    \\ \prod_{j\leq N-1}
    \left[N_{\omega,j}^{\rm (in)} e^{i\omega t}+(1+N_{\omega,j}^{\rm (out)})e^{-i\omega t}\right]\bigg\}\, .\label{eq:P(E)-gen}
\end{multline}
Here $S=E_{\mathrm c}^{-(N-1)}{N^N}/{(N-1)!}$, and $E_{\mathrm c}=e^2/2C$
is the Coulomb charging energy of a single junction ($C$
is a single junction capacitance) and for the Cooper pair transport $e\to 2e$.
Equations~\eqref{eq:cotunneling_rate} and \eqref{eq:P(E)-gen} were
derived in a first order in tunneling Hamiltonian.
Shown in Fig.~\ref{fig:cotunneling} is a diagrammatic representation
of Eq.\,\eqref{eq:P(E)-gen} for $N=3$.

A generalization of the results obtained for a single junction including
the structure of the collision integral and the concept of the effective
temperature Eq.\,\eqref{eq:T_eff}, onto large arrays is straightforward.
As long as temperatures are not extremely low~\cite{Lopatin1}, the
charge transfer in large arrays is dominated by
the inelastic cotunneling and the cascade energy relaxation.
The tunneling carriers generate e-h pairs~\cite{Lopatin1,Lopatin2} serving
as an environment exchanging the energy with the tunneling current and then
slowly losing it to the bath.

\section{Cascade two-stage relaxation: general applications}
In the preceding sections we have formulated a general approach to description
of electronic transport in mesoscopic tunnel junctions mediated by the energy
exchange between the charge carriers and the environment, which, in its turn,
relaxes the energy to thermostat.
In what follows we will show that there is a rich variety of the
seemingly disparate kinetic phenomena in
tunnel junction arrays that allow for a natural and transparent
description within the unique framework of
this hierarchical or cascade relaxation concept.
We illustrate the power of our approach applying it to phenomena
of overheating~\cite{heatingReviewA,heatingReviewB},
the Coulomb anomaly~\cite{STM1D,Grabert_Devoret}, the transport in the granular
systems governed by the cotunneling processes~\cite{Efros,Grabert_Devoret}
and mediated by the electron-hole environment.
The concept of the cascade two-stage energy relaxation appears to be
of the fundamental importance to revealing the microscopic mechanism
of the insulator-to-superinsulator transition and the nature of low temperature transport
in the superinsulating state~\cite{FVB,FVBlp,VinNature}.
We show, finally, that the concept of the cascade energy relaxation
resolves the long standing puzzle of temperature-independent pre-exponential
factor in variable range hopping
conductivity~\cite{Khondaker1999,Shlimak1999,Ghosh,Yakimov,Baturina2007,Baturina2008a,Baturina2008b,EpsilonExp4}.
We demonstrate that the underlying physical mechanism behind all the above is the
imbalance between the intense energy exchange of the charge carriers
with the (nonequilibrium) environment
and the comparatively weak coupling of the environment to the phonon bath.

\subsection{Nonequilibrium Coulomb anomaly}
\begin{figure}[h]
\center
  \includegraphics[width=9.0cm]{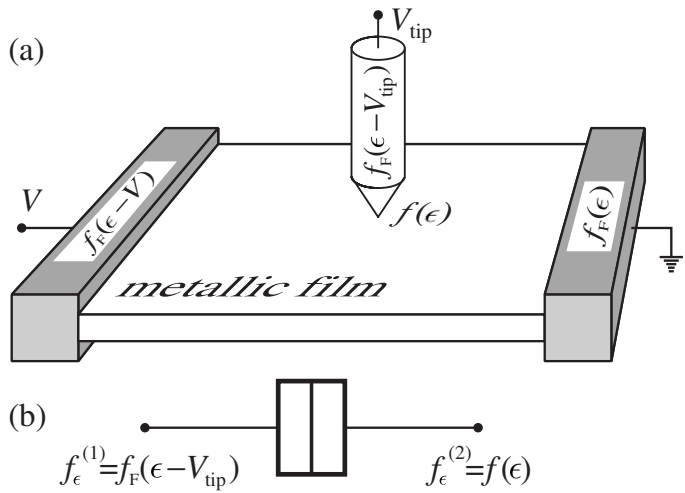}
  \caption{ (a) The quasi-2D disordered metallic film attached at the edges to two electrodes.
The tunneling density of states (TDOS) can be determined
using the transport measurement done with
the help of the Scanning Tunnel Microscope (STM) tip.
(b) Equivalent circuit that describes transport between the STM-tip and metal
(in nonequilibrium state).
}
\label{fig:wire}
\end{figure}

Consider the quasi-2D disordered
metallic film attached at the edges to two electrodes, see Fig.\,\ref{fig:wire}(a).
The Coulomb anomaly arises due to electromagnetic fluctuations
associated with the electron-electron interaction in the wire
(electron-hole environment excitation)
and manifest itself as the suppression of the local tunnel density of states (TDOS)
at small energies~\cite{Aronov-Altshuler}.
Experimentally the density of states can be determined by means
of tunneling transport measurements with the scanning tunnel microscope (STM)
tip serving as one of the electrodes, see
Fig.\,\ref{fig:wire}(a) and Refs.\,\cite{STM1D,Giazotto}.
The corresponding equivalent circuit is shown in Fig.\ref{fig:wire}(b).
The resistor characterizes the contact between the tip and the quasi-2D metallic film.
This is the same resistor that appears in Fig.\,\ref{fig:e-ph_int}.

In the absence of the current, i.e. in an equilibrium, the distribution function
of electrons at any point
of the film is the Fermi function.
Then the current voltage characteristics of the junction,
$I(V_{\rm tip})$, where $V_{\rm tip}$ is the potential of the tip,
can be found following the recipes of Refs.\,\cite{Devoret_basic,Aronov-Altshuler}:
we have to use Eqs.\,\eqref{eq:current}-\eqref{eq:S_begin_total} with
the \textit{equilibrium} environment distribution function and take $\rho(\omega)
=2\Imm\int_{\mathbf q}\tilde U/(D q^2-i\omega)^{2}$, where $\tilde U$
is the dynamically-screened Coulomb interaction in the metal and
$D$ is the diffusion coefficient.
Then from the differential conductance,
$dI/dV_{\rm tip}$, we get the local density of states
(the same as in Ref.\cite{Kamenev-Andreev})
and find all the standard Coulomb anomaly features,
see Refs.\,\cite{Aronov-Altshuler,Kamenev-Andreev}.

As the current starts to flow through the metallic film,
then the distribution function of electrons, $f(\epsilon,\mathbf{r})$,
at low temperatures (where the phonon bath is frozen out) becomes nonequilibrium
and should be found from the kinetic equation~\cite{STM1D,Giazotto},
$\triangle f=0$ with the edge conditions relating electron distribution
function in the metal with the electron distribution functions at the reservoirs.
The solution of the kinetic equation at the center of the metallic film
can be approximated as follows:
\begin{gather}\label{eq:f}
    f(\epsilon)=\frac12[f_{\scriptscriptstyle{\mathrm F}}(\epsilon)+f_{\scriptscriptstyle{\mathrm F}}(\epsilon-V)]\,,
\end{gather}
where $V$ is the electrical potential of the left electrode that pushes
the current through the metal, see Fig.\,\ref{fig:wire}(a).
Here it was assumed that the electron inelastic scattering length, $l_{\rm in}$, is
larger than the separation between the electrodes attached to the metallic film.
The electron-hole environmental modes in the metallic film, which are responsible for
the Coulomb anomaly, become now nonequilibrium.
Since the phonon bath is frozen, then $\tau_{\rm env-bath}\gg\tau_{\rm e-env}$, so the
stationary distribution function $N_\omega$ is to be found from Eq.\,\eqref{eq:N_e-env}.
Then we obtain,
\begin{gather}\label{eq:NN}
    N_{\omega}=n_{\omega},\qquad
    n_\omega=(2\omega)^{-1}\int d\epsilon f_{\epsilon_+}(1-f_{\epsilon_-})\,.
\end{gather}
We calculate the  tunneling rate between for electrons passing from the tip to
the metallic film using Eq.\,\eqref{eq:S_begin_total}:
\begin{gather}\label{eq:S_begin_total_new}
        \overrightarrow{\Gamma}=\frac1{R_{\scriptscriptstyle {\mathrm T}}}
        \int_{\epsilon\epsilon'}f_\epsilon^{(\mathrm{tip})}
        (1-f_{\epsilon'})P^<(\epsilon-\epsilon')\, ,
   \end{gather}
where $f_\epsilon^{(\mathrm{tip})}=f_F(\epsilon-V_{\mathrm{tip}})$
and $f_\epsilon$ is given in Eq.\,\eqref{eq:f}.
Note that when evaluating $P^{<}$, we used the nonequilibrium
environmental bosonic distribution function $N_\omega$ as given by Eq.\,\eqref{eq:NN}.
Equations~\eqref{eq:current} and \eqref{eq:S_begin_total_new} yield the current,
flowing through the tip, $I(V_{\rm tip})$, from which one determines the
 TDOS as:
\begin{gather}\label{eq:tdos}
    \nu_{T}(\epsilon)\propto \frac{dI(V_{\rm tip})}{dV_{\mathrm{tip}}}\biggl|_{V_{\mathrm{tip}}\to\epsilon}\,.
\end{gather}
One finds after some algebra that the TDOS acquires
the `superstructure' on the energy scale of the order of $T_{\rm eff}(V)$:
two dips develop at the energies corresponding to the positions of the
Fermi level in the tip and in the film,
where the electronic
distribution function experiences discontinuous jumps (i.e. at $\epsilon=0$
and $\epsilon=V$)~\cite{STM1D,Mirlin}.

\subsection{Negative differential resistance and overheating}

If the rate of the energy supply from the external bias to the charge carriers
exceeds the rate of energy losses to the environment, the phenomenon of \textit{overheating}
takes place and the energy distribution function of the current carriers
noticeably deviates from the equilibrium distribution
function~\cite{heatingReviewA,heatingReviewB}.
One of the characteristic manifestations of the overheating effect is the onset
of the ``falling'' region of the $I$-$V$ curve where the
differential conductivity  $G={\partial I}/{\partial V}<0$.
The corresponding $I(V)$ characteristics is referred to as
that of the $S$-type if the current is the multi-valued function of the voltage,
and the $I$-$V$ curve of the $N$-type corresponding to the case
where the current is nonmonotonic but still remains a single valued function of the voltage.
The phenomenon of overheating has been a subject of the incremental interest and
extensive studies during the decades,
see the detailed Volkov and Kogan review~\cite{heatingReviewA}, and the impressive
progress in understanding of the underlying mechanisms was achieved.
Recently, for instance, the
ideas of~\cite{heatingReviewA} were applied to
description of nonlinear low temperature
current-voltage characteristics in disordered granular metal~\cite{Shuler}.
The concept of the two-stage relaxation of the present work enables us to construct
a general approach to overheating taking into the consideration also non-phonon
mechanisms of relaxation. A general scheme is as follows.  Let us consider the
far-from-equilibrium tunneling transport mediated by an environment.  Let the
environmental excitations \textit{emitted} by tunneling carriers be
``hot'' (``up" lines in Figs.\,\ref{fig:e-ph_int}-\ref{fig:cotunneling})
whereas the environmental excitations absorbed by tunneling carriers are ``cold''
(``down" lines in Figs.\,\ref{fig:e-ph_int}-\ref{fig:cotunneling}).
This implies that there exist two distinct energy scales characterizing the
frequency distributions,
$N_\omega^{\rm(in)}$ and $N_\omega^{\rm(out)}$, corresponding to
 emitted and absorbed environment excitations, respectively:
\begin{gather}\label{1221}
    T_{\rm eff}^{\rm (out)} =\lim _{\omega\to 0} \omega N_\omega^{\rm(out)},
    \\
    T_{\rm eff}^{\rm (in)} = \lim _{\omega\to 0} \omega N_\omega^{\rm(in)}.
\end{gather}
These two temperature scales are, in general, different provided the effective inelastic length,
$L_{\rm \phi}$, describing thermalization of the excitations with the phonon bath (thermostat)
is of the same order or smaller than the hot region where tunneling electrons strongly
interact with the environment.
Then the absorbed environmental excitations come from the outside of the hot region
and have the temperature of the bath, $T_{\rm eff}^{\rm (in)}\approx T_{\rm bath}$.
The emitted excitations are hot, and their effective temperature
is the function of the applied voltage (current).
The temperature of the hot excitations is to be found from the heat
balance equation following the recipes of Refs.~\cite{heatingReviewA,heatingReviewB}.
Namely, it is obtained by integrating the product of the kinetic equation for electron
distribution function [see, e.g., Eq.\,\eqref{eq:sc}] and the electron energy $\epsilon$,
over the volume and energy:
\begin{gather}\label{9}
    P\backsimeq\frac{\mathcal{E}(T_{\rm eff}^{\rm (in)})}{\tau_{\rm e-env}(T_{\rm eff}^{\rm (in)})}-\frac{\mathcal{E}(T_{\rm eff}^{\rm (out)})}{\tau_{\rm e-env}(T_{\rm eff}^{\rm (out)})},
\end{gather}
where $\tau_{\rm e-env}(T)$ is the temperature dependent electron-environment
inelastic scattering rate, $\mathcal{E}(T)$ is of the order of total the energy of electrons
having the temperature $T$ in the volume involved
 and $P\backsimeq V^2/R(T_{\rm eff}^{\rm (out)})$ is
the (Joule) heat produced by the passing current.
The $I$-$V$ characteristics comes out as solutions to Eq.\,\eqref{9} with the
specific relations between $\mathcal{E}$ and the applied voltage and current.
The resulting $I$-$V$ curves may come up as the so called
$S$- or $N$-type of the $I$-$V$
characteristics depending on the specific temperature dependencies
$R(T_{\rm eff}^{\rm (out)})$, see Ref.~\cite{heatingReviewA}.
Making use of Eq.\,\eqref{eq:sc} we can estimate $\tau_{\rm e-env}(T)$.
It is proportional to the coefficient standing
in the scattering integral by $f^{(1)}_{\epsilon_1}$:
\begin{gather}
\begin{split}
\frac1{\tau_{\rm e-env}}\sim &\int d\omega\nu_\omega \nu_2d\epsilon_2\left(\frac{\rho}
{\omega\nu_\omega}\right)W_{12}\times
\\
\biggl\{
&\delta(\epsilon_{12}-\omega)[(N_\omega+1)
(1-f^{(2)}_{\epsilon_2})+N_\omega f^{(2)}_{\epsilon_2}]+
\\
&\delta(\epsilon_{12}+\omega)[N_\omega
(1-f^{(2)}_{\epsilon_2})+(N_\omega+1) f^{(2)}_{\epsilon_2}]\biggr\}\,.\label{eq:heating}
\end{split}
\end{gather}

The equation\,\eqref{eq:heating} is derived for a single tunnel junctions.
To describe the overheating effects in arrays of many junctions one needs
an appropriate generalization of a single junction approach.
An important example of such a multi-junction system is an array of metallic granules
in the insulating state.
In granular metals the role of environment is taken by the electron-hole pairs.
In this case $ \tau_{\rm e-env}$
is electron inelastic relaxation time, $T_{\rm eff}^{\rm (in)}=T$ is the bath temperature.
The energy scale $\mathcal{E}$ can be roughly approximated by the energy of the quasiparticles
in the bulk of the sample of the volume $\Omega$ as $\mathcal{E}(T)\sim \pi^2\nu \Omega T^2/6$.
Employing the standard theory of the tunneling conductivity
in the granular materials ~\cite{Lopatin2} one then can arrive at
the $S$-type $I$-$V$-characteristics for granular metals.

\subsection{Electron-hole environment in a disordered metal}

To gain an insight into the behavior of the electron-hole environment in mesoscopic
tunnel junctions and reveal the role of Coulomb interactions, which are
instrumental to formation of the environment properties,
let us employ the Finkel'stein's theory of Fermi-liquid
fluctuations in a disordered granular metal~\cite{Fin}.  The theory treats the
Coulomb potential as the contact interaction.
This offers a pretty good description of disordered metals, where Coulomb potential is
well screened to $\delta$-function.  Of course, in the close vicinity of the disorder-driven
metal-insulator (or superconductor-insulator) transition where the real
Coulomb fields come into play, the orthodox Finkel'stein's theory is not valid,
and one needs the approach capable to accommodate the long-range Coulomb effects.

We adopt the Keldysh representation which is most adequate for discussing dynamic effects.
The action describing Fermi-liquid fluctuations acquires the form:
\begin{gather}
iS_{\scriptscriptstyle{\mathrm F}}=
-\frac{1}{4}\sigma_{\scriptscriptstyle{\mathrm N}}
\tr\left[(\check{\partial}_{\mathrm r}Q)^2-\frac{4}{D}\check{\partial}_{\mathrm t}Q\right]
-\frac{i\Gamma_{\rho}}{4\nu}
\int_{tx}\{(\hat{\rho}_1)_{tx}(\hat{\rho}_2)_{tx}
+2i\nu\tr[\overrightarrow{\phi}^{\tau}\sigma_{\mathrm x}\overrightarrow{\phi}]\}.
\label{Finaction}
\end{gather}
Here $\hat{\rho}_{1(2)}$ is the operator of the Fermion fluctuations density
in the classical (quantum) sector,
\begin{equation}
\hat{\rho}_{1}=-\frac{2\pi\nu}{1+F_{\rho}}\left[\tr(\sigma_{\mathrm x}Q_{{\mathrm t},{\mathrm t'}})+\frac{\phi_1}{2\pi}\right]\,,\,\,\,\,\hat{\rho}_{2}=-\frac{2\pi\nu}{1+F_{\rho}}\left[\tr(\sigma_{\mathrm x}Q_{{\mathrm t},{\mathrm t'}})+\frac{\phi_2}{2\pi}\right],
\end{equation}
$\overrightarrow{\phi}$ are the conjugated fields,
$\Gamma_{\rho}=F_{\rho}/(1+F_{\rho})$, $F_{\rho}$ is
the contact amplitude modeling the screened Coulomb interaction in the singlet channel.
The Fourier transform of the typical retarded fluctuation propagator
describing fluctuations has the form:
\begin{gather}
{\cal D}^{{\cal{\scriptscriptstyle{\mathrm R}}}}=\frac{1}{Dq^2-i\omega(1+\Gamma_{\rho})}\,,
\end{gather}
where $\omega$ are the Matsubara frequencies.
The diffusion poles of the  propagators thus possess the structure,
$Dq^2-i\omega$, $Dq^2-i\omega(1+\Gamma_{\rho})$. One sees that
this description applies and the spectrum of the excitations
is stable only as long as the single granule mean level spacing
remains the smallest energy parameter, i.e. $\delta<Dq^2, \omega, \omega\Gamma_{\rho}$.
This implies the development of the excitations spectrum instability and opening
the energy gap at temperatures
\begin{equation}
  T<T^*=\frac{\delta}{\Gamma_{\rho}}\,.\label{manybody}
\end{equation}
Formation of the energy gap in the spectrum of the environmental excitations
is a phenomenon that tremendously influences the tunneling electronic transport.
One sees straightforwardly that as the energy gap, $T^*$, in the spectrum of
electron-hole excitations appears at $T<T^*$, the spectral probability $\rho(\omega)$
for the electron-environment interaction
vanishes in the interval $0<\omega<T^*$. Then Eq.\,(\ref{eq:cotunneling_rate})
yields the suppression of the tunneling current to the practically zero magnitude.
The physical significance and meaning of this result is merely that
as the gap in the excitation spectrum opens, the environment ceases to
efficiently mediate the energy relaxation from the tunneling carriers,
impeding thus the tunneling current.

Interestingly, the similar temperature that marked vanishing of the
conductivity by the e-e interactions in the absence of
coupling of electronic system to phonons in the model of disordered
quantum wire was found in
Refs.~\cite{Gornyi2005,BAA2006} by reducing the electron
conductivity to the Anderson model on the Bethe lattice.  This
suppression of conductivity was interpreted as the Anderson localization in Fock
space.  We reiterate here that models based on the contact e-e interactions hold only
as long as the long-range Coulomb effects are effectively screened.
As a result its applicability, for example, to the description
of physical phenomena in the vicinity of metal-insulator or
superconductor-insulator transition needs special justification.
In the next section we propose an approach capable to explicitly
account for the long-range Coulomb interactions in the
critical region of disorder-driven superconductor-insulator transition.

%%%%%%%%%%%%%%%%%%%%%%%%%%%%%
%%%%%%%%%%%%%%%%%%%%%%%%%%%%%
%%%%%%%%%%%%%%%%%%%%%%%%%%%%%

\subsection{Superinsulating behavior}

Let us consider charge transfer in a two-dimensional array of superconducting tunnel junctions
(or, equivalently, Josephson junction network) in the insulating state, i.e. under the
conditions that $E_{\scriptscriptstyle{\mathrm J}}<E_{\mathrm c}$,
where $E_{\scriptscriptstyle{\mathrm J}}$ is the Josephson coupling energy,
and $E_{\mathrm c}$ is the charging energy related to the capacitance $C$  between the two
adjacent granules (or the capacitance of a single Josephson junction).
We focus on the limit $C\gg C_0$, where $C_0$ is the capacitance of a
single junction to the ground.
The electric properties of the array are quantified by the screening length
$\lambda=a\sqrt{C/C_0}$, where $a$ is the size of the elemental unit of
the Josephson junction network.
At distances  $R<\lambda$, the electric charges interact according the logarithmic law,
the energy of the interaction being $\propto\ln(R/\lambda)$,
at larger distances Coulomb interaction is exponentially screened.
If the linear dimension of an array, $L$ does not exceed $\lambda$, the charge
plasma, comprising the environment in the superconducting array,
experiences the charge binding-unbinding Berezinskii-Kosterlitz-Thouless
(BKT) transition~\cite{BKTa,BKTb,BKTc,BKTd,Mooij1990,Fazio1991} at
$T=T_{\scriptscriptstyle{\mathrm{BKT}}}\simeq E_{\mathrm c}$~\cite{Mooij1990,Fazio1991}.
In the low temperature BKT phase positive and negative charges are bound into the dipole
pairs, whereas above the transition the charges form a free plasma.
Binding positive and negative charges into pairs implies that at $T\simeq E_{\mathrm c}$
the energy gap $\simeq E_{\mathrm c}$ opens
in the spectrum of unbound charges.  As we have discussed in the preceding
section, opening the energy gap in the environmental excitation spectrum
results in vanishing $\rho(\omega)$ in the interval $0<\omega<E_{\mathrm c}$.

The low temperature transport in the large JJA arrays was discussed 
in~\cite{FVB,FVBlp,VinNaturer} within the framework of the Feynman-Vernon influence 
functional technique.  
The charge transfer is realized by the dipole excitations that unbind and further 
tunnel under the applied bias.  
In the course of tunneling the propagating charges generate and leave behind 
the strings of the dipole (or electron-hole) excitations that accommodate 
the charges energy difference at different superconducting islands.  
As shown in~\cite{FVB,FVBlp,VinNature} the system of dipoles excited 
at Josephson junctions can be mapped onto the ensemble of quantum rotators.  
The subsequent averaging of the current with respect to all the states 
of the rotator ensemble is equivalent to averaging over the thermodynamic 
bath, the role of which is thus taken by the dipole environment.  
Thus the relaxation process ensuring the tunneling current is governed 
by the dipole environment and the subsequent phonon emission by these dipole excitations.

Now we discuss the same process from the viewpoint of the nonequilibrium 
approach developed in the present work.
Following further the line of reasoning used in preceeding sections 
and employing again Eq.\,(\ref{eq:cotunneling_rate}),
we arrive at the conclusion that
 at $T<T_{\scriptscriptstyle{\mathrm{BKT}}}$,
the charge transfer in the array of the superconducting tunnel junctions is suppressed.
Remembering now, that at low temperatures tunneling current in an array of
superconducting junctions is governed by the
cotunneling processes~\cite{Lopatin1,Lopatin2} described in Section 1.3.
Analyzing the contribution from higher orders into the
cotunneling process, one finds that the current suppression holds in all orders.

On a qualitative level the effect of the gap opening  in the charge plasma excitation spectrum
on the charge transfer in the superconducting tunneling array can be described as follows.
Starting with Eq.~\eqref{eq:current} and using
Eqs.~\eqref{eq:cotunneling_rate}-\eqref{eq:P(E)-gen}, one can find
    \begin{equation}
        I\propto\exp(-E/W)\,,
        \label{eq:SI-current}
    \end{equation}
where $E$ is the characteristic energy for the charge transfer between the granules.
In the system with the linear size $L<\lambda$,
$E=E_{\mathrm c}\ln(L/a)$, where $E_{\mathrm c}=2e^2/C$ for the Cooper pair,
and $E_{\mathrm c}=e^2/2C$ for the quasiparticle.
The quantity $W$ is the energy scale associated with the rate of the energy exchange between
the tunneling charges and the environment.
While the rigorous derivation for $W$ is not available at this point,
the estimates suggest the interpolation formula
        \begin{gather}
             W\simeq\frac{E_{\mathrm c}}{\exp(E_{\mathrm c}/T)-1}\,.
             \label{eq:W}
        \end{gather}
The meaning of this formula is that the relevant energy scale characterizing
the tunneling rate is the energy gap that enters with the weight equal to
the Bose-kind filling factor describing the probability of exciting the unbound charges.
Well above the charge BKT transition Eq.\,\eqref{eq:W} gives $W=T$ (this reflects the fact that
the number of the unbound charges
is determined by the equipartition theorem and is proportional
to $T/E_{\mathrm c}$).  This yields
    \begin{gather}
        I\propto\exp[-E_{\mathrm c}\ln(L/a)/T]\,, \,\,\,T\gtrsim T_{\scriptscriptstyle{\mathrm{BKT}}}\,.
        \label{eq:highTcurrent}
    \end{gather}
One has to bear in mind that this formula holds only at temperatures
not too high above the charge BKT transition, where Coulomb interactions
are not completely screened.
At $T\gg T_{\scriptscriptstyle{\mathrm{BKT}}}$, $E=E_{\mathrm c}\ln(\lambda/a)$
and as the screening length becomes of the order of $a$, $E\simeq E_{\mathrm c}$.
In this temperature region the system exhibits the `bad metal' behavior.
Notably, Eq.\,\eqref{eq:highTcurrent} looks like a formula for the thermally activated
current, however one has to remember that the physical mechanism behind
the considered charge transfer is quantum mechanical tunneling process which can take
place only if the mechanisms for the energy relaxation are switched on.

At low temperatures, $T\ll T_{\scriptscriptstyle{\mathrm{BKT}}}\simeq E_{\mathrm c}$,
the characteristic energy from Eq.\,\eqref{eq:W} is $W=E_{\mathrm c}\exp(-E_{\mathrm c}/T)$.
The unbound charges that have to mediate the energy relaxation from the
tunneling carriers are in an exponentially short supply, and one finds
    \begin{gather}
        I\propto\exp[-\ln(L/a)\exp(E_{\mathrm c}/T)]\,, \,\,\,T\ll T_{\scriptscriptstyle{\mathrm{BKT}}}\,,
        \label{eq:lowTcurrent}
    \end{gather}
reproducing the results of~\cite{FVB,FVBlp,VinNature} for the Cooper pair conductivity.
The additional suppression of the electronic transport
in Josephson junction arrays as compared to the activation regime was indeed experimentally
observed in Refs.~\cite{Kanda,Yamaguchi}.

The outlined picture of the tunneling current in the superconducting junction array
applies to thin superconducting films close to superconductor-insulator
transition (SIT)~\cite{Baturina2007,Baturina2008a,Baturina2008b}.
Indeed, in the close vicinity of the SIT, the disorder-induced spatial modulations
of the superconducting gap give rise to the electronic phase separation
in a form of the texture of weakly coupled superconducting
islands~\cite{STM_TiN} with the characteristic spatial scale of the
order of several superconducting coherence lengths $\xi$.
The consequences of this phase separation are two-fold.
First, the fine balance between disorder and Coulomb forces near
the SIT results, on one hand, in the effective reduction of
the disorder strength, and, on the other hand, in the emergence of the
strong non-screened Coulomb field in the spaces between the islands.
This field can be argued to reconstruct the island texture into
a nearly regular array of weakly coupled superconducting islands, i.e. into
an array of mesoscopic superconducting junctions.
Note that the regular lattice of the superconducting islands can be also
induced by nonlocal elastic fields due to coupling between
the film and the substrate~\cite{islands}.
Second, in the critical region near the SIT this island array is on
the verge of the percolation-like transition between the superconductor
and insulator (the cluster of the islands touching each other and
traversing the sample means the emergence of global superconductivity).
Since in such two-phase systems the dielectric constant diverges
on approach the
percolation transition~\cite{Dubrov,Efros1976,EpsilonExp1,EpsilonExp2,EpsilonExp3,
EpsilonExp4}, the thin films in the critical region of the SIT develop a
huge \textit{global} dielectric permeability.
This means that on the distances $L<\varepsilon d$, where $d$ is the film thickness,
all the electric fields, emerging due to fluctuations in the local charge of the 2D
environment, remain trapped within the film, and, therefore, the charge environment
is nothing but the 2D neutral plasma, where charges interact with each other
according to the logarithmic law.  Therefore the 2D charge environment
experiences the binding-unbinding charge BKT transition
at $T=T_{\scriptscriptstyle{\mathrm{BKT}}}$,
and the energy gap opens in the environmental excitations spectrum at
$T\leq T_{\scriptscriptstyle{\mathrm{BKT}}}$, giving rise to
suppression of \textit{both Cooper pair- and normal quasiparticle currents},
since both tunneling processes -- irrespectively to whether it is transfer
of Cooper pairs or quasiparticles across the junction -- can occur only if the
energy exchange with an environment is possible.
Thus opening the gap in an environmental spectrum due to long-range Coulomb effects
and the resulting suppression of the
tunneling current offers a microscopic mechanism behind the
insulator-to-superinsulator transition and the conductivity in the
superinsulating state~\cite{FVB,FVBlp,VinNature}.

Note in this connection, that in the recent work~\cite{Syzr}
the dc conductivity of an array of Josephson junctions in the insulating state is discussed
in terms of transport of Cooper pairs through the narrow band formed by
the Josephson coupling and distorted by \textit{weak} disorder.  
This model for the Cooper pair transfer is in a sense complimentary to that
of the present work where we discuss the opposite limit
of sequential tunneling between the granules with the essentially different tunneling levels.

\subsection{Variable Range Hopping conductivity}

The notion of the cascade two-stage relaxation is a key to resolving the controversy of
the variable range hopping (VRH) conductivity.
Many experimental studies
of the hopping conductivity in doped
semiconductors~\cite{Khondaker1999,Shlimak1999,Ghosh}, in arrays of
quantum dots~\cite{Yakimov}, and in
disordered superconducting films~\cite{Baturina2007,Baturina2008a,Baturina2008b,EpsilonExp4},
revealed the temperature independent pre-exponential
factor, indicating the
non-phonon mechanism of relaxation.
What more, it had the universal values
of the integer multiples of $e^2/h$.
According to the early ideas by Fleishman, Licciardello, and Anderson~\cite{Fleishman},
the universal pre-exponential factor may evidence that the energy relaxation
is due to electron-electron (e-e) rather than the electron-phonon interactions.
On the other hand, according to Refs.~\cite{Gornyi2005,BAA2006} the e-e relaxation alone
(i.e. in the absence of the coupling to phonons, which are in any case
present in a real physical system)
cannot ensure a finite conductivity below the
certain localization temperature.
Namely, in the absence of phonons the system of interacting electrons subject to disorder
undergoes a localization transition (`many-body localization') in Fock space.
The concept of the cascade energy relaxation developed in the present work
resolves this controversy.
The first stage is the fast energy exchange
between the hopping electrons
and the electron-hole environment (electrons do not see sparse phonons).
Since this process involves only
electron-electron interactions, the resulting pre-exponential factor
does not depend on temperature.
The second-stage process is the transfer of the energy from the electron-hole environment to
the phonon thermostat.
This process which is of course necessary to ensure the current,
does not influence the pre-exponential factor.

\section{Conclusions}

In conclusion, we have developed a theory
of the far from the equilibrium tunneling transport
in arrays of tunnel junctions in the limit $1/\tau_{\rm{e-env}}\gg 1/\tau_{\rm env-bath}$.
We have demonstrated that the  energy relaxation ensuring the
low-temperature tunneling current occurs as a cascade two-stage process:
the tunneling charges lose their energy to an intermediate agent, environment,
and the latter relaxes the energy further to the thermostat.
We have derived the tunneling $I$-$V$ characteristics and shown that
they are highly sensitive to the
structure of the spectrum of the excitations of the environmental modes.
In particular, opening the energy gap in the excitation spectrum suppresses the
tunneling current.
As an important example we have considered a two-dimensional array of the
superconducting junctions where the two-stage relaxation occurs via the two dimensional
Coulomb plasma of positive and negative charges with logarithmic interaction, which experiences
the charge binding-unbinding BKT transition at $T=T_{\scriptscriptstyle{\mathrm{BKT}}}$.
We argued that the gap in the plasma excitation spectrum that emerges
at $T<T_{\scriptscriptstyle{\mathrm{BKT}}}$ gives rise to suppression of both,
tunneling Cooper pair- and quasiparticle currents, thus
offering the possible microscopic mechanism for the insulator-to-superinsulator transition and
the low temperature transport in the superinsulating state.
We considered  applications of our
general approach to several physical systems and low temperature transport phenomena,
including Coulomb anomaly in 2D disordered metals, overheating effects, electron-hole
environment in disordered metals, and the origin of temperature independent pre-exponential
factor in hopping conductivity.

\section{Acknowledgments}
We are grateful to R. Fazio, A. Shytov, A. Gurevich, A. Chubukov,
I. Burmistrov, and Ya. Rodionov for useful discussions.
This work was supported by the U.S. Department of Energy Office of Science
under the Contract No. DE-AC02-06CH11357,
by the Programs of the Russian Academy of Sciences,
and by the Russian Foundation for Basic Research
(Grant Nos. 09-02-01205 and 09-02-12206).

\printindex

\begin{thebibliography}{99}
\bibitem{footnote111} In an equilibrium, $q_{\scriptscriptstyle{\mathrm T}}$
defines the long wave cut-off of the diffusive fluctuations.
In can be shown that in the nonequilibrium case the same cut
off retains with the change $T\to T_{\rm eff}$~\cite{SVB}.
\bibitem{env_DoS}
In an LC-line environment the excitations spectrum is linear,
$\omega=v k$, where $v=c/\sqrt {LC}$, so $\nu_\omega=1/v$.
\end{thebibliography}
\end{document}